# Improved Accuracy and Precision In Simultaneous Myocardial $T_1$ and $T_2$ mapping with Multi-Parametric SASHA (mSASHA)

Kelvin Chow[1], Genevieve Hayes[2], Jacqueline A Flewitt[2], Patricia Feuchter[2], Carmen Lydell[2], Andrew Howarth[2], Joseph J Pagano[3], Richard B Thompson[4], Peter Kellman[5], James A White[2]

[1]Cardiovascular MR R&D, Siemens Medical Solutions USA, Inc., Chicago IL, USA

[2]Stephenson Cardiac Imaging Centre, University of Calgary, Calgary AB, Canada

[3]Division of Pediatric Cardiology, University of Alberta, Edmonton AB, Canada

[4]Department of Biomedical Engineering, University of Alberta, Edmonton AB, Canada

[5]National Heart, Lung, and Blood Institute, National Institutes of Health, Bethesda MD, USA

*This manuscript has been submitted to Magnetic Resonance in Medicine and is under peer review*

**Running title**: Myocardial $T_1$ and $T_2$ Mapping with mSASHA

**Data availability statement**: The data that support the findings of this study are available from the corresponding author, KC, upon reasonable request.

**Corresponding author**:

Dr. Kelvin Chow, Cardiovascular MR R&D, Siemens Medical Solutions USA, Inc., Chicago IL, USA

Email: kelvin.chow@siemens-healthineers.com

**Abstract**

*Purpose*: To develop and validate a multi-parametric SAturation-recovery single-SHot Acquisition (mSASHA) cardiac $T_1$ and $T_2$ mapping technique with high accuracy and precision in a single breath-hold.

*Methods*: The mSASHA acquisition consists of 9 images in an 11 heartbeat breath-hold -- the first without preparation, 6 images with saturation recovery preparation, and 2 images with both saturation recovery and $T_2$-preparation. $T_1$ and $T_2$ values were calculated using a 3-parameter model. mSASHA was validated in simulations and phantoms on a Siemens 3T Prisma scanner with comparison to a joint $T_1$-$T_2$ technique with a 4-parameter model. mSASHA values were compared to reference MOLLI, SASHA and $T_2$p-bSSFP sequences in 10 healthy volunteers.

*Results*: mSASHA had high accuracy compared to reference spin-echo measurements, with an average of -0.7±0.4% $T_1$ error and -1.3±1.3% $T_2$ error. mSASHA coefficient of variation (CoV) in phantoms for $T_1$ was lower than MOLLI (0.7±0.1% vs 0.9±0.2%, p<0.01) and similar to reference $T_2$p-bSSFP for $T_2$ (1.4±0.6% vs 1.5±0.5%, p>0.05). In simulations, 3-parameter mSASHA fitting had higher precision than 4-parameter joint $T_1$-$T_2$ fitting for both $T_1$ and $T_2$. In-vivo myocardial mSASHA $T_1$ was similar to conventional SASHA (1523±18 ms vs 1520±18 ms, p>0.05) with similar CoV to both MOLLI and SASHA (3.3±0.6% vs 3.1±0.6% and 3.3±0.5% respectively, p>0.05 for both). Myocardial mSASHA $T_2$ values were 37.1±1.1 ms with similar precision to $T_2$p-bSSFP (6.7±1.7% vs 6.0±1.6%, p>0.05).

*Conclusion*: mSASHA provides high accuracy cardiac $T_1$ and $T_2$ quantification in a single breath-hold, with similar precision to reference MOLLI and linear $T_2$p-bSSFP reference techniques.

**Background**

Cardiac MR is commonly used for the assessment of myocardial disease by measuring abnormalities in tissue relaxation times (1). Increased native longitudinal ($T_1$) relaxation times have been correlated with myocardial fibrosis (2) and can be combined with post-contrast $T_1$ imaging to calculate the extracellular volume fraction (ECV) (3-5), a physiologically relevant metric of fibrosis. Increased transverse ($T_2$) relaxation times have been associated with myocardial edema and inflammation (6). Using conventional, non-parametric, $T_1$- and $T_2$-weighted imaging techniques, regional myocardial disease is subjectively or semi-quantitatively characterized by relative differences in signal intensity versus reference tissue, the latter assumed to be healthy. However, this poses challenges for the detection of global tissue injury, and inherently limits objective comparisons between individuals and serial evaluations of the same subject. Parametric $T_1$ and $T_2$ mapping techniques are intrinsically quantitative, objectively calculating relaxation times in each voxel, eliminating the need for reference tissue comparisons. These address previous challenges of weighted imaging, and optimization of quantitative techniques are focused on increasing imaging efficiency with high accuracy and precision.

$T_1$ mapping is being increasingly adopted in clinical practice as an quantitative, complementary marker of extracellular matrix expansion, commonly interpreted in parallel with conventional late gadolinium enhancement (LGE) imaging. Two dominant techniques currently exist – the MOdified Look-Locker Inversion recovery (MOLLI) sequence (7,8) and the SAturation-recovery single-SHot Acquisition (SASHA) sequence (9). The MOLLI sequence gained widespread adoption due in part to its high precision and the multitude of studies correlating its $T_1$ and derived ECV values to various metrics of disease severity in a wide range of cardiomyopathies (4,10-13). However, MOLLI is known to significantly underestimate the true $T_1$ value, with a bias from physiological parameters such as heart rate (7,14,15), magnetization transfer (16), $T_1$ (17), $T_2$ (17) and non-physiological parameters such as inversion pulse efficiency (18), $B_0$ (19), and $B_1$ (15) inhomogeneities. By comparison, the SAturation-recovery single-SHot Acquisition (SASHA) sequence (9) is generally regarded as highly accurate with robustness to confounders such as magnetic field inhomogeneities ($B_0$ and $B_1$) (15,20), heart rate (9), $T_1$ (9), $T_2$ (9), and magnetization transfer effects (16). However, the original SASHA sequence used a 3-parameter model with lower precision compared to MOLLI due to the reduced dynamic range of saturation recovery approaches compared to inversion recovery techniques (15). Recent work has improved the precision of SASHA using a combination of optimized sampling times (21) and a 2-parameter model with a variable flip angle (VFA) readout (22).

$T_2$ mapping is commonly performed using a combination of a $T_2$ preparation ($T_2$p) pulse and a balanced steady-state free precession (bSSFP) readout with linear k-space ordering (6). This approach is appealing due to its simplicity and high signal-to-noise ratio (SNR), but is known to overestimate $T_2$ values due to $T_1$ recovery effects during the imaging readout (6). Centric k-space ordering improves accuracy, but has reduced SNR and also image blurring associated with centric ordering (6).

Several recently developed techniques have combined $T_1$ and $T_2$ mapping into a single sequence, including cardiac magnetic resonance fingerprinting (cMRF) (23,24), multitasking (25), joint $T_1$/$T_2$ mapping with saturation recovery (26,27), and 3D-QALAS (28). These techniques are attractive because $T_1$ and $T_2$ provide complimentary information about tissue microstructure and are often acquired in the same study. Combined multi-parametric techniques also have the unique benefit of being inherently co-registered, avoiding concerns about spatial misalignment when combining data from separate acquisitions. The joint $T_1$-$T_2$ mapping proposed by Akçakaya et al. (26), is a saturation-recovery based sequence similar to SASHA with additional images having $T_1$ and $T_2$ preparation. This technique demonstrated good accuracy in phantoms, but had moderate precision due to a 4-parameter fitting model.

In this study, we propose a novel multi-parametric SASHA (mSASHA) sequence that combines $T_1$ and $T_2$ mapping in a single saturation-recovery based sequence with improved precision using a 3-parameter model (29). The sequence is validated in simulations and phantoms and then evaluated in healthy subjects alongside the conventional SASHA, MOLLI, and $T_2$p-bSSFP sequences.

**Methods**

*Theory*

The multi-parametric SASHA (mSASHA) acquisition consists of a series of single-shot balanced steady-state free-precession (bSSFP) images, starting with an image without magnetization preparation, followed by a set of saturation-recovery (SR) images, and finally a set of images with both saturation and $T_2$-preparation (Fig. 1a). A similar acquisition with combined SR+$T_2$p images has been previously described by Akçakaya et al. (26) and Guo et al. (27). The number of SR images and SR+$T_2$p images is flexible, with 6 SR images and 2 SR+$T_2$p images used in this study for an 11 heartbeat (HB) breath-hold. An additional recovery heartbeat is used between the saturation pulse and $T_2$p pulses was used to increase the SNR of the SR+$T_2$p images.

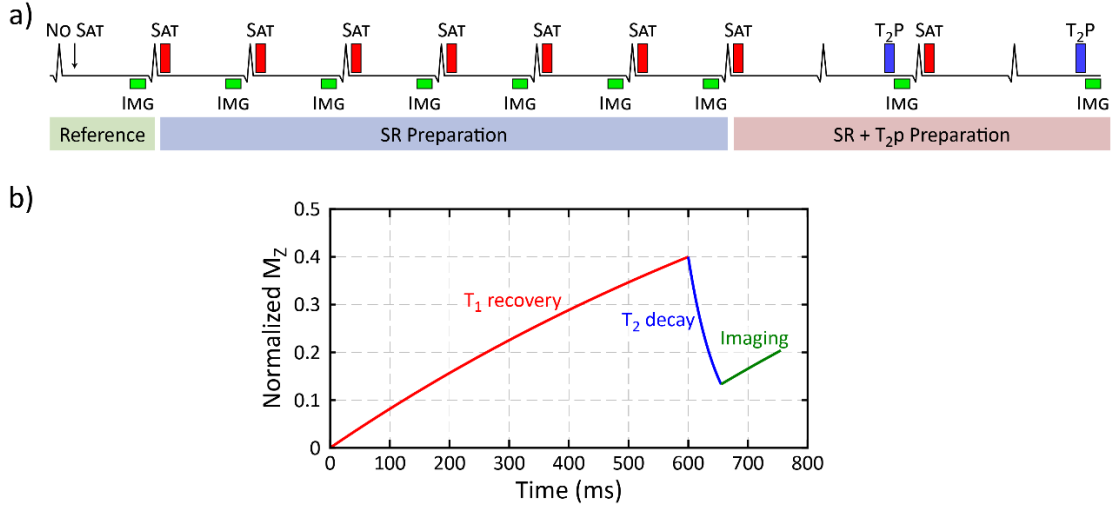

**Fig. 1. a)** Sequence diagram for multi-parametric SASHA (mSASHA), consisting of a non-prepared reference image, a series of saturation-recovery prepared images, and a series of saturation-recovery and $T_2$-prepared images. Imaging is performed using single-shot balanced steady-state free-precession readouts. **b)** Signal model for mSASHA, where the signal intensity for each image is the combination of $T_1$ recovery, optional $T_2$ decay, and the effect of the imaging readout.

The longitudinal magnetization for each image can be described as a combination of $T_1$ weighting from the saturation pulse, optional $T_2$ weighting from the $T_2$-preparation pulse, and an additional effect from the linear bSSFP readout (Fig. 1b). Reference images are a limiting case of full $T_1$ recovery and no $T_2$ decay and images with only SR preparation have no $T_2$ decay component.

In the joint $T_1$-$T_2$ model described by Akçakaya, this can be represented as a 4-parameter equation (Eq. 1) (26):

$$S(T_1, T_2, A, b) = A\left\{\left(1 - e^{-TS/T_1}\right) e^{-TE/T_2}\right\} + b \qquad \text{Eq. 1}$$

Where:

- TS is the saturation recovery time between the end of the saturation pulse and the start of the $T_2$p pulse for SR+$T_2$p images or start of imaging for SR only images,
- TE is the $T_2$-preparation duration,
- $T_1$ is the longitudinal recovery time,
- $T_2$ is the transverse recovery time, and
- A and b are constants.

In this 4-parameter model, the "b" parameter is a constant offset used to account for the effect of the imaging readout. During a bSSFP readout, the evolution of the longitudinal magnetization can be approximated as exponential recovery with an apparent relaxation rate of (Eq. 2) (30):

$$R_1' = R_1 \cos^2(\theta/2) + R_2 \sin^2(\theta/2) \qquad \text{Eq. 2}$$

Where:

- $R_1'$ is the apparent relaxation rate
- $R_1$ is the true longitudinal relaxation rate, equal to $1/T_1$
- $R_2$ is the transverse relaxation rate, equal to $1/T_2$
- $\theta$ is the bSSFP readout flip angle

Using a partial sinusoidal variable flip angle (VFA) readout, the flip angles prior to the center of k-space are relatively small and $R_1'$ is approximately equal to $R_1$. In this case, the signal model can be simplified to 3-parameter one (Eq. 3):

$$S(T_1, T_2, A) = A\left\{1 - \left[1 - \left(1 - e^{-TS/T_1}\right)e^{-TE/T_2}\right]e^{-TD/T_1}\right\} \qquad \text{Eq. 3}$$

where TD is the known fixed duration between the end of the $T_2$ preparation pulse and the center of k-space and all other variables are defined as for Eq. 1 and Eq. 2. Data from all images are included during model fitting, where TE=0 is used for images without $T_2$ preparation.

This 3-parameter (3p) model is hypothesized to improve the precision of calculated $T_1/T_2$ values compared to the 4-parameter (4p) model due to the simplification of the model. The accuracy and precision of 3p mSASHA is validated with comparison to the 4p joint $T_1$-$T_2$ model in simulations and also to reference standard spin echo in phantom experiments.

*Phantom Imaging*

Accuracy and precision of the prototype mSASHA sequence was evaluated in a $NiCl_2$-doped agarose phantom with physiologic $T_1/T_2$ combinations representative of myocardium and blood in various physiological and pathological states (T1MES, Resonance Health, Perth, Australia) (31). Imaging was performed on a 3T MAGNETOM Prisma scanner (Siemens Healthcare, Erlangen, Germany). An inversion-recovery spin-echo (IR-SE) acquisition was used for reference $T_1$ measurements with sequence parameters: 360×124 mm$^2$ field of view, 8 mm slice thickness, 192×66 matrix size, turbo factor 1, echo time (TE) 12 ms, repetition time (TR) 10,000 ms, with 16 inversion times (TI) between 100 and 5,000 ms. Additional spin-echo images were used for reference $T_2$ measurements with 6 TEs between 12 and 200 ms in separate acquisitions, no inversion, and all other parameters matched.

Conventional cardiac $T_1$ mapping was also performed in the phantom using the prototype MOLLI and SASHA sequences using a simulated heart rate of 60 bpm and sequence parameters detailed in Table 1. Briefly, MOLLI was acquired with a 5(3)3 sampling scheme and both a typical 35° and a 20° flip angle to reduce sensitivity to $B_0/B_1$ inhomogeneity. SASHA was acquired with a variable flip angle (VFA) readout with a 70° maximum (22) and a numerically optimized saturation pulse train, providing <1% residual Mz over a wide range of $B_0/B_1$ (20). Saturation recovery times (TS) for all SR images were set to 600 ms or maximum allowed by heart rate to improve precision (21). Conventional cardiac $T_2$ mapping was performed using $T_2$-prepared images with both linear bSSFP and centric gradient recalled echo (GRE) readouts in separate measurements. Linear $T_2$p-bSSFP protocol parameters are detailed in Table 1 and centric $T_2$p-GRE images were acquired with a 192×120 matrix size, 83% phase resolution, 1.35/3.20 ms TE/TR, 15° flip angle and all other parameters matched to $T_2$p-bSSFP.

The proposed mSASHA sequence was acquired with one non-prepared anchor image, 6 SR images with 600 ms TS, and 2 SR+$T_2$p images with 1600 ms TS and 55 ms $T_2$p. An optimized $T_2$-preparation module was used consisting of adiabatic half-passage tan/tanh tip-down and tip-up pulses and 3 adiabatic full-passage tan/tanh refocusing pulses. A 100° maximum VFA readout was used with an additional high-contrast image to improve motion correction (32) and other parameters in Table 1.

As the mSASHA protocol had only two SR+$T_2$p images with identical $T_2$p times, the 4-parameter joint $T_1$-$T_2$ model was underdetermined and fitting was not possible because only 3 unique TS/$T_2$p combinations were available. In order to evaluate the precision of 4p joint $T_1$-$T_2$ fitting in comparison with mSASHA, a separate 13-heartbeat "variable $T_2$p" mSASHA dataset was acquired, substituting one of the SR+$T_2$p images with a 30 ms $T_2$p duration and adding an SR image with one recovery heartbeat (1600 ms TS).

| | Multi-parametric SASHA | SASHA | MOLLI | Linear $T_2$-prepared bSSFP |
|---|---|---|---|---|
| **Field of View** | (340-380)×(210-324) mm² | | | |
| **Slice Thickness** | 8 mm | | | |
| **Matrix Size** | 256×(118-186) | | | |
| **Phase Resolution** | 75% | | | |
| **Partial Fourier** | 7/8th | | | |
| **Echo Time** | 1.28-1.31 ms | 1.23-1.29 ms | 1.09-1.15 ms | 1.36-1.42 ms |
| **Repetition Time** | 2.91-3.04 ms | 2.87-3.00 ms | 2.64-2.77 ms | 3.09-3.23 ms |
| **Flip Angle** | 100° maximum variable flip angle | 70° maximum variable flip angle | 20° or 35° constant flip angle (5 linear ramp up pulses) | 70° constant flip angle (5 linear ramp up pulse) |
| **Parallel Imaging** | R=2, 36 separately acquired reference lines | | | |
| **Breath-hold duration** | 11 heartbeats | 11 heartbeats | 11 heartbeats | 9 heartbeats |
| **Sampling Scheme** | - 1 non-prepared<br>- 6 SR (600 ms TS, or maximum allowed by heart rate)<br>- 2 SR+$T_2$p (600+RR ms TS, 55 ms $T_2$p) | - 1 non-prepared<br>- 10 SR (600 ms TS, or maximum allowed by heart rate) | - 5(3)3 scheme<br>- 100 ms minimum TI<br>- 80 ms TI increment | - 0, 30, 55 ms $T_2$p duration, separated by 3 recovery heartbeats |

**Table 1**. In-vivo protocol parameters for mSASHA, SASHA, MOLLI, and linear $T_2$-prepared bSSFP sequences for $T_1$ and $T_2$ mapping.

*Simulations*

Monte-Carlo simulations were performed to assess the precision of all fitting models, including the proposed 3-parameter mSASHA model compared to the 4-parameter joint $T_1$-$T_2$ model. In order to account for imaging effects on the signal intensity, measured signal intensities from the 13 HB variable $T_2$p acquisition in the phantoms were used as input for simulations. Simulations were performed for SNR values between 50 and 200 in steps of 25, with 10,000 repeats at each SNR. In each repeat, noise was added to produce a Rician distribution with ν being the signal intensities and σ being 1/SNR. Data was fit to 3p mSASHA and 4p joint $T_1$-$T_2$ models. Data from only the reference and SR images (i.e. excluding SR+$T_2$p images) were also fit to 2-parameter and 3-parameter SASHA models ($T_1$ only) for comparison.

*In-Vivo Imaging*

Ten healthy volunteers with no known history of cardiovascular disease were recruited with written informed consent and imaged on a Siemens 3T MAGNETOM Prisma scanner. mSASHA was acquired in a mid short-axis slice with 6 SR images having a 600 ms TS or maximum allowed by heart rate and both SR+$T_2$p images having a TS as above with one recovery heartbeat and a 55 ms $T_2$p. Additional sequence parameters are described in Table 1. MOLLI (both 20° and 35° flip angles), SASHA, and linear $T_2$p-bSSFP were acquired as reference techniques with parameters as described above. Based on phantom and simulation results, the 13-heartbeat variable $T_2$p protocol was not acquired in-vivo.

*Image Analysis*

All images were reconstructed with matched parameters in Gadgetron (33) to facilitate comparison of precision while parametric maps were calculated in MATLAB (R2020a, The MathWorks, Natick, USA). IR-SE $T_1$ values were calculated using a 3-parameter exponential recovery model and SASHA $T_1$ values were calculated using a 2-parameter exponential recovery model assuming ideal saturation. MOLLI $T_1$ values were calculated using a 3-parameter model with Look-Locker correction and an inversion efficiency correction factor of 1.0365 (18). Spin-echo, centric $T_2$p-GRE, and linear $T_2$p-bSSFP $T_2$ values were calculated using a 2-parameter exponential decay model. mSASHA $T_1$/$T_2$ values were calculated using the proposed 3-parameter model (Eq. 2). Joint $T_1$/$T_2$ values were calculated using 4p joint $T_1$-$T_2$ model (Eq. 1) for the variable $T_2$p data in phantoms, but not the standard mSASHA protocol as the model was underdetermined for these TS/$T_2$p times. Coefficient of variation (CoV) was calculated for each phantom vial as the standard deviation of $T_1$ or $T_2$ over the vial divided by the mean.

In-vivo images from all sequences were motion corrected using the Advanced Normalization Tools (ANTs) software (34), with high-contrast images used to improve motion correction for mSASHA (32). Regions of interest (ROIs) were manually contoured for individual phantom vials, entire left ventricular myocardium, and left ventricular blood pool. $T_1$ and $T_2$ values are reported as mean ± standard deviation across the entire ROI. Paired two-sided Student's t-test were used to test for statistical significance with p<0.05.

**Results**

*Phantoms*

Measured $T_1$ and $T_2$ values from all sequences are reported in Tables 2 and 3 respectively and $T_1$ and $T_2$ maps are shown in Fig. 2. The proposed mSASHA with 3-parameter model had excellent $T_1$ accuracy of -0.7±0.4% and $T_2$ accuracy of -1.3±1.3% as compared to spin-echo measurements. Accuracy for the 13-heartbeat variable $T_2$p protocol was similar for 3p and 4p models in both $T_1$ (-0.7±0.4% vs 0.2±0.9%, p>0.05) and $T_2$ fitting (-2.1±1.1% vs 3.1±1.5%, p>0.05).

Conventional SASHA had high $T_1$ accuracy (-0.9±0.2%) while MOLLI underestimated $T_1$ values in vials with myocardial-like short $T_2$, with increased underestimation at 35° compared to 20°, up to 11.0% error. Centric $T_2$p-GRE had good accuracy for most vials except vial 6, where the long $T_1$ value representative of native blood likely resulted in incomplete magnetization recovery during the recovery interval between images and significant $T_2$ underestimation. Blurring in the phase encode (vertical) direction is visible in the centric $T_2$p-GRE map (Fig. 2), consistent with the centric k-space ordering. Linear $T_2$p-bSSFP had significant error for all vials, particularly the ones with post-contrast myocardial-like $T_1/T_2$ combinations, with up to 78% error.

Parametric map precision in the 13-heartbeat variable $T_2$p data was significantly improved using the 3p model compared to the 4p model for both $T_1$ CoV (0.6±0.2% vs. 2.3±1.3%, p<0.01, Table 2) and $T_2$ CoV (1.5±0.7% vs. 3.2±1.4%, p<0.01, Table 3). $T_1$ CoV with 3p mSASHA (0.68±0.14%) was similar MOLLI 35° (0.70±0.11%, p>0.05) and lower than both MOLLI 20° (0.92±0.23%, p<0.01) and SASHA (0.76±0.13%, p<0.05). $T_2$ CoV with 3p mSASHA was not statistically significantly different compared to linear $T_2$p-bSSFP (1.4±0.6% vs 1.5±0.5%, p>0.05). $T_2$ precision was not compared with the centric $T_2$p-GRE acquisition due to the difference in spatial resolution.

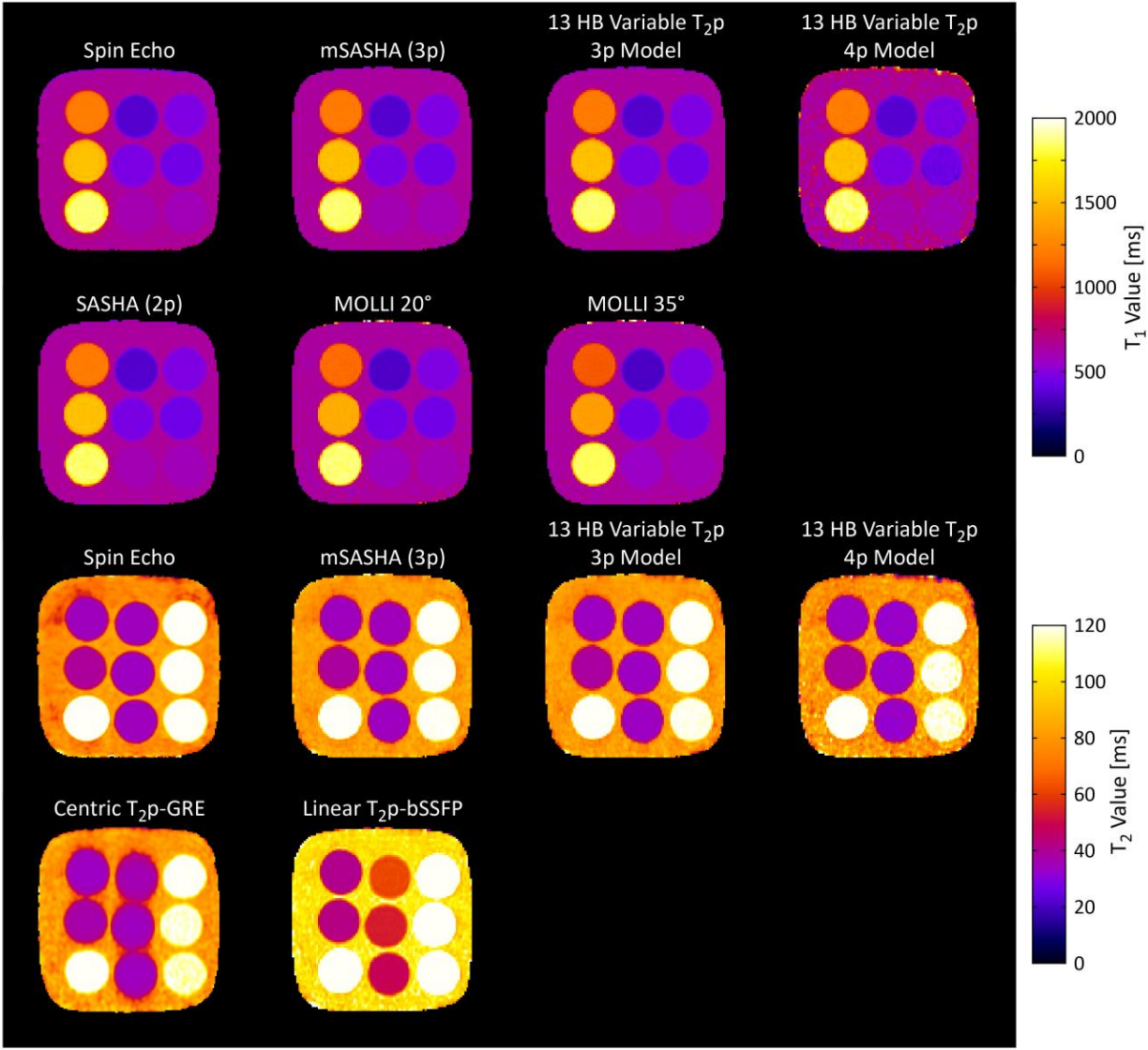

**Fig. 2**. $T_1$ and $T_2$ maps from mSASHA and various reference sequences in a T1MES phantom.

|  |  |  | mSASHA | 13 Heartbeat Variable $T_2p$ |  |  |  |
|---|---|---|---|---|---|---|---|
|  | Vial | Spin Echo | 3p Model | 3p Model | 4p Model | SASHA | MOLLI 20° | MOLLI 35° |
| Myocardial-Like | 1 | 1216.7 ± 4.4 | 1202.8 ± 8.1 | 1202.9 ± 7.9 | 1211.5 ± 11.4 | 1205.7 ± 8.4 | 1137.0 ± 7.6 | 1083.3 ± 7.4 |
| Myocardial-Like | 2 | 1527.5 ± 9.2 | 1512.9 ± 13.2 | 1513.0 ± 12.7 | 1516.8 ± 19.8 | 1514.6 ± 13.4 | 1429.9 ± 10.7 | 1360.9 ± 9.8 |
| Myocardial-Like | 3 | 606.4 ± 2.9 | 601.3 ± 4.1 | 600.9 ± 4.2 | 614.2 ± 13.9 | 601.6 ± 4.9 | 571.4 ± 7.3 | 553.9 ± 5.2 |
| Myocardial-Like | 4 | 470.9 ± 2.2 | 465.8 ± 3.5 | 465.6 ± 3.1 | 473.1 ± 7.7 | 465.9 ± 3.7 | 442.2 ± 5.3 | 434.1 ± 2.9 |
| Myocardial-Like | 5 | 365.1 ± 2.0 | 361.2 ± 2.4 | 361.1 ± 2.1 | 367.6 ± 4.5 | 360.7 ± 2.6 | 344.1 ± 3.0 | 340.0 ± 2.3 |
| Blood-Like | 6 | 1856.7 ± 10.9 | 1850.5 ± 15.6 | 1850.7 ± 14.9 | 1852.3 ± 30.2 | 1842.5 ± 18.5 | 1855.4 ± 15.9 | 1829.5 ± 10.7 |
| Blood-Like | 7 | 590.7 ± 1.4 | 590.5 ± 3.9 | 590.6 ± 3.5 | 592.7 ± 21.0 | 587.2 ± 4.2 | 594.1 ± 6.4 | 590.9 ± 4.5 |
| Blood-Like | 8 | 441.8 ± 1.4 | 439.7 ± 2.6 | 440.3 ± 2.2 | 436.3 ± 21.5 | 437.9 ± 3.2 | 441.3 ± 4.2 | 441.1 ± 3.2 |
| Blood-Like | 9 | 483.1 ± 1.1 | 479.7 ± 1.9 | 480.3 ± 1.7 | 475.4 ± 13.5 | 478.6 ± 2.5 | 482.8 ± 3.1 | 482.2 ± 2.6 |

**Table 2**. $T_1$ measurements with various sequences in NiCl$_2$ doped agarose phantom. Values are reported as mean ± standard deviation across each vial. Coefficients of variation (CoV) were calculated as standard deviations divided by means.

|  |  |  | mSASHA | 13 Heartbeat Variable $T_2p$ |  | Centric | Linear |
|---|---|---|---|---|---|---|---|
|  | Vial | Spin Echo | 3p Model | 3p Model | 4p Model | $T_2p$-GRE | $T_2p$-bSSFP |
| Myocardial-Like | 1 | 34.6 ± 0.2 | 34.4 ± 0.3 | 34.1 ± 0.3 | 33.9 ± 0.5 | 34.0 ± 0.3 | 40.8 ± 0.3 |
| Myocardial-Like | 2 | 39.3 ± 0.2 | 38.9 ± 0.5 | 38.6 ± 0.4 | 38.5 ± 0.5 | 37.1 ± 0.4 | 42.2 ± 0.4 |
| Myocardial-Like | 3 | 34.6 ± 0.4 | 34.2 ± 0.5 | 33.9 ± 0.4 | 32.8 ± 1.2 | 34.3 ± 0.4 | 48.6 ± 0.6 |
| Myocardial-Like | 4 | 34.1 ± 0.2 | 34.2 ± 0.4 | 33.8 ± 0.3 | 32.8 ± 1.0 | 34.5 ± 0.5 | 53.8 ± 0.6 |
| Myocardial-Like | 5 | 34.6 ± 0.3 | 34.9 ± 0.3 | 34.5 ± 0.3 | 33.0 ± 1.0 | 36.1 ± 0.4 | 61.5 ± 0.6 |
| Blood-Like | 6 | 203.7 ± 1.4 | 198.4 ± 5.1 | 198.4 ± 6.1 | 198.5 ± 6.5 | 132.6 ± 3.1 | 172.4 ± 3.5 |
| Blood-Like | 7 | 125.0 ± 0.9 | 120.9 ± 1.7 | 119.8 ± 1.8 | 119.6 ± 3.9 | 115.6 ± 2.4 | 142.4 ± 2.8 |
| Blood-Like | 8 | 127.1 ± 0.9 | 124.3 ± 1.4 | 123.6 ± 1.4 | 125.6 ± 7.7 | 118.2 ± 1.8 | 155.0 ± 2.6 |
| Blood-Like | 9 | 129.8 ± 2.9 | 127.2 ± 2.7 | 126.3 ± 2.8 | 127.8 ± 5.1 | 123.4 ± 1.9 | 155.8 ± 3.5 |

**Table 3**. $T_2$ measurements with various sequences in NiCl$_2$ doped agarose phantom. Values are reported as mean ± standard deviation across each vial. Coefficients of variation (CoV) were calculated as standard deviations divided by means.

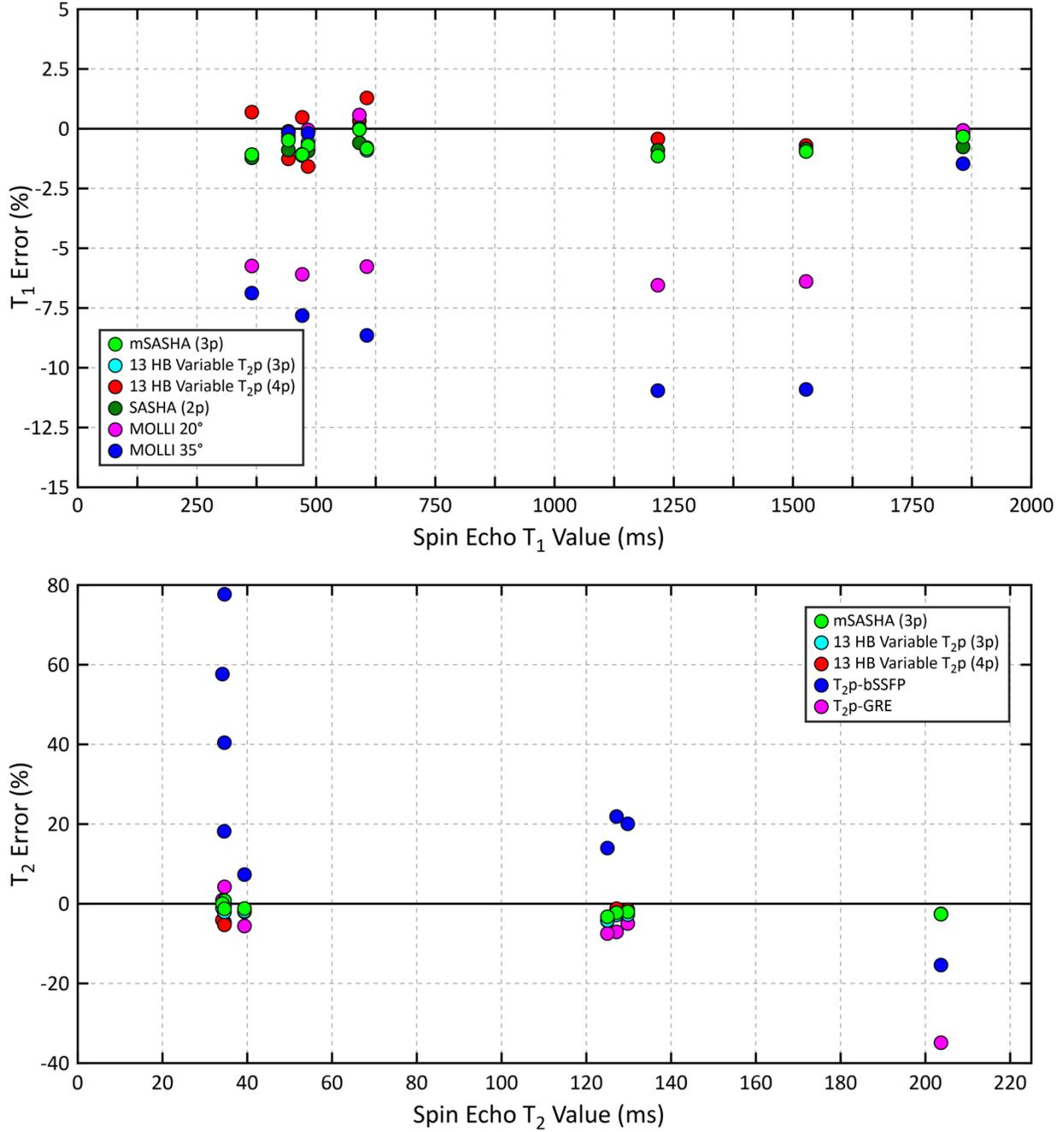

**Fig. 3**. Accuracy of mSASHA $T_1/T_2$ and various reference sequences compared to gold standard spin-echo measurements. Accuracy for 13 HB variable $T_2p$ with 3p fitting is very similar to mSASHA (3p) and plotted data points are largely obscured.

*Simulations*

Monte-Carlo simulations of precision (Fig. 4) were consistent with phantom data. In the native myocardium (vial 1) and blood (vial 6) simulations, the $T_1$ IQR of 4p joint $T_1$-$T_2$ was similar to 3p SASHA as previously described (26) and $T_1$ IQR of 3p mSASHA was similar to 2p SASHA. In both cases, $T_1$ precision was not adversely affected by the use of combined $T_1/T_2$ models compared to $T_1$-only fitting models. The $T_1$ IQR at an SNR of 100 for 4p joint $T_1$-$T_2$ was 2.6× and 2.4x worse than 3p mSASHA for native myocardium and blood, respectively. Both 4p joint $T_1$-$T_2$ and 3p SASHA had significantly higher IQR compared to 3p mSASHA when $T_1$ was short or $T_2$ was long due to ill conditioning of these models with the TS/$T_2$p times used. Overall, 3p mSASHA had more robust precision over a range of $T_1/T_2$ values.

$T_2$ precision with 3p mSASHA was superior to 4p joint $T_1$-$T_2$ in all cases (Fig. 4), with $T_2$ IQR with 4p joint $T_1$-$T_2$ being 1.7× larger for native myocardium and 1.1x larger for native blood. However, $T_2$ IQR with 4p joint $T_1$-$T_2$ was substantially worse than 3p mSASHA in simulations with short $T_1$ representing post-contrast imaging.

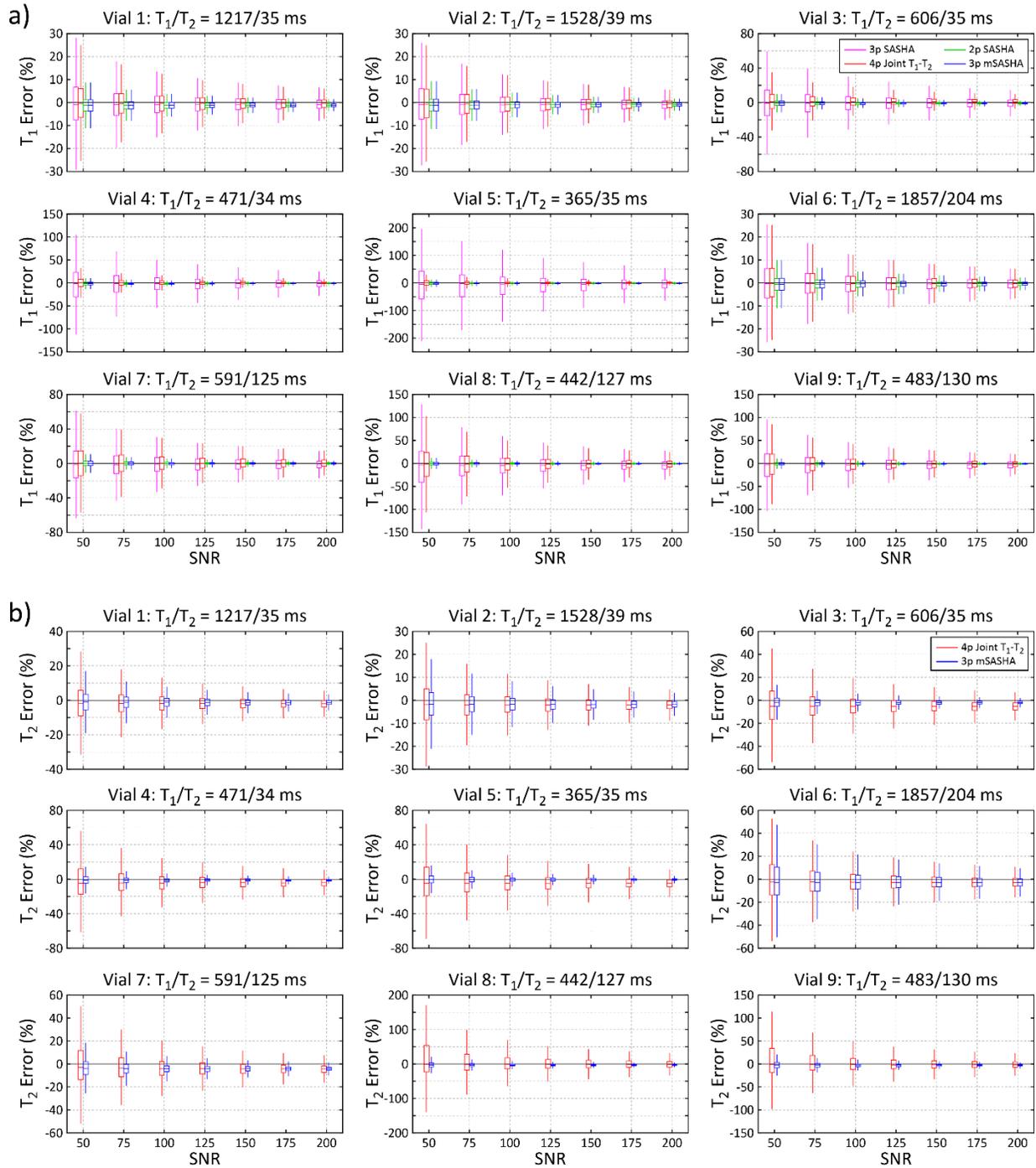

**Fig. 4.** Monte Carlo simulation results of **a)** $T_1$ and **b)** $T_2$ precision with various signal models. Boxes indicate the first and third quartiles and whiskers indicate 1.5×IQR.

*In-Vivo*

Volunteers were an average of 44.2±13.1 years old (7 female) with a heart rate of 69±12 bpm during imaging. MOLLI 20° data from one subject was discarded due to operator error (insufficient recovery period prior to start of acquisition). $T_1$ and $T_2$ values from all sequences are detailed in Table 4 and $T_1$ and $T_2$ maps from all sequences in a healthy volunteer are shown in Fig. 5. mSASHA $T_1$ values were not statistically significantly different from SASHA $T_1$ values for both the myocardium (1523±18 ms vs 1520±18, p>0.05) and left ventricular blood pool (2054±61 ms vs 2060±65 ms, p>0.05). MOLLI $T_1$ values were underestimated in the myocardium compared to SASHA, with greater underestimation at 35° than 20° (1218±23 ms vs 1275±24 ms, p<0.001). $T_2$ values with linear $T_2$p-bSSFP were higher than mSASHA in the myocardium (41.3±1.7 ms vs 36.7±1.1 ms, p<0.001) but lower in the blood (127.9±9.8 ms vs 160.8±14.8 ms, p<0.001), consistent with over/underestimation observed in phantoms for similar $T_1/T_2$ combinations.

The CoV for mSASHA $T_1$ values was not statistically significantly different than SASHA in both the myocardium and blood pool (p>0.05, Table 4). Myocardial mSASHA $T_1$ CoV (3.3±0.6%) was not statistically significantly different than either MOLLI 20° (3.1±0.6%) or MOLLI 35° (3.3±0.8%) (p>0.05 for both). However, mSASHA $T_1$ CoV in the blood (3.0±0.7%) was larger than both MOLLI 20° (2.0±0.4%) and MOLLI 35° (1.6±0.3) (p<0.001 for both). $T_2$ CoV was not statistically significantly different between mSASHA and linear $T_2$p-bSSFP in both the myocardium (6.7±1.0% vs 6.0±1.6%, p>0.05) and the blood (9.8±4.0% vs 9.1±3.1%, p>0.05).

|  | T$_1$ Values | | | | T$_2$ Values | |
|---|---|---|---|---|---|---|
|  | mSASHA | SASHA | MOLLI 20° | MOLLI 35° | mSASHA | Linear T$_2$p-bSSFP |
| Myocardial value (ms) | 1523.1 ± 17.8 | 1519.5 ± 18.0 | 1275.0 ± 23.9 | 1217.7 ± 22.9 | 36.7 ± 1.1 | 41.3 ± 1.7 |
| Blood value (ms) | 2054.1 ± 60.9 | 2060.1 ± 65.4 | 1872.7 ± 71.4 | 1855.7 ± 64.1 | 160.8 ± 14.8 | 127.9 ± 9.8 |
| Myocardial SD (ms) | 50.7 ± 9.4 | 49.6 ± 7.2 | 40.0 ± 7.4 | 40.7 ± 9.5 | 2.5 ± 0.6 | 2.5 ± 0.7 |
| Blood SD (ms) | 62.7 ± 14.9 | 64.5 ± 16.0 | 37.5 ± 7.6 | 29.4 ± 5.6 | 16.0 ± 7.4 | 11.7 ± 4.3 |
| Myocardial CoV (%) | 3.3 ± 0.6 | 3.3 ± 0.5 | 3.1 ± 0.6 | 3.3 ± 0.8 | 6.7 ± 1.7 | 6.0 ± 1.6 |
| Blood CoV (%) | 3.0 ± 0.7 | 3.1 ± 0.8 | 2.0 ± 0.4 | 1.6 ± 0.3 | 9.8 ± 4.0 | 9.1 ± 3.1 |

**Table 4**. T$_1$ and T$_2$ measurements in 10 healthy volunteers with mSASHA, SASHA, MOLLI, and linear T$_2$p-bSSFP.

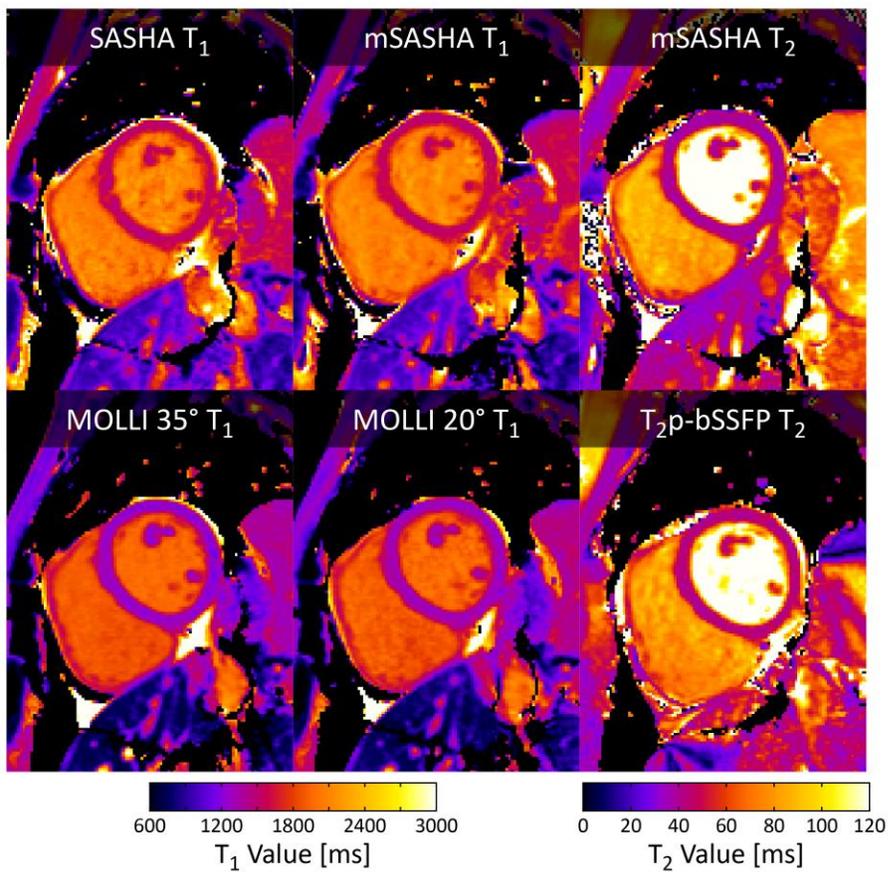

**Fig. 5**. T$_1$ and T$_2$ maps from mSASHA, SASHA, MOLLI, and linear T$_2$p-bSSFP sequences in a healthy subject at 3T.

**Discussion**

We demonstrated mSASHA to be a robust and clinically translatable technique that maintains the superior accuracy and precision of the SASHA $T_1$ mapping sequence while providing simultaneous $T_2$ quantification in a single breath-hold. Use of a variable flip angle readout enabled formulation of a 3-parameter model for simultaneous $T_1$ and $T_2$ mapping with multi-parametric SASHA, significantly improving precision compared to the 4-parameter joint $T_1$-$T_2$ model (26,27). Overall, mSASHA resulted in the lowest $T_1$ and $T_2$ error (<1.5%) in phantoms and maintained in-vivo precision similar to that provided by the most commonly used MOLLI $T_1$ and linear $T_2$p-bSSFP clinical sequences.

Simultaneous and accurate quantification of $T_1$ and $T_2$ enables improved characterization of microstructural changes encountered with myocardial disease. The widely used MOLLI $T_1$ mapping sequence is recognized to be confounded by alterations in $T_2$ (15,17), while linear $T_2$p-bSSFP values are conversely confounded by $T_1$ (6). In contrast, mSASHA provided accurate quantification of both $T_1$ and $T_2$ across a wide range of physiologically relevant $T_1$/$T_2$ combinations, enabling each parameter to be reliably measured even if the other is altered. The intrinsic co-registration of mSASHA $T_1$ and $T_2$ maps may be particularly useful in the assessment of the area-at-risk, as fibrosis and edema-mediated microstructural changes may provide unique influences on these respective parameters. Simultaneous and accurate measurements may also permit improved understanding of the relationship between $T_1$ and $T_2$ changes across a variety of myocardial disease states, including of inflammatory myocarditis, iron overload, and genetically-mediated cardiomyopathies.

*Comparisons with Other Parametric Mapping Sequences*

Several alternative sequences exist for simultaneous cardiac $T_1$ and $T_2$ quantification. The joint $T_1$ and $T_2$ mapping sequence proposed by Akçakaya et al. (26) has several similarities to mSASHA as they both use a similar single-shot bSSFP acquisition with variable saturation and $T_2$-preparation modules. However, the reduction in the number of fit parameters with 3p mSASHA compared to 4p joint $T_1$-$T_2$ significantly improved precision in both Monte Carlo simulations and phantom data. $T_1$ precision of 4p joint $T_1$-$T_2$ was comparable to $T_1$-only 3-parameter SASHA fitting in simulations, consistent with reports by Akçakaya et al. and Guo et al. (26,27). The improved precision of 2p SASHA versus 3p SASHA is also consistent with previous literature (15,22). Precision for 4p joint $T_1$-$T_2$ and 3p SASHA was poor for short $T_1$ and long $T_2$ combinations due to poor sampling of the recovery curve. Precision could potentially be improved for these models through optimization of the TS/$T_2$p sampling times (21,35) for different expected $T_1$/$T_2$ values, such as different sampling schemes for native and post-contrast. In contrast, precision of 3p mSASHA was stable across simulations in all phantom vials.

With respect to in-vivo $T_1$ values obtained using joint versus separate acquisition sequences, Akçakaya et al., previously found similar mean values between joint $T_1$ and SASHA $T_1$ in-vivo at 1.5T (26). However, a recent multi-slice extension of the joint $T_1$-$T_2$ approach by Guo et al. at 3T showed significant underestimation of $T_1$ values compared to SASHA (191 ms in volunteers, 111 ms in patients), with unknown causes (27). In our study, we observed no underestimation of in-vivo mSASHA $T_1$ values vs SASHA (1523±18 ms vs 1520±18 ms, p>0.05). These values were also similar to literature values for SASHA at 3T, with Weingärtner et al. measuring 1523±41 ms in 20 volunteers (36), Guo et al. reporting 1521±17 ms in the basal slice of 13 healthy volunteers (27), and Heidenreich et al. describing 1460±67 ms in remote myocardium of 19 patients (37). Agreement between mSASHA and SASHA in both simulations and previously published in-vivo SASHA values suggests that reference normal values from SASHA could potentially be used for mSASHA as well. However, further comparison in a larger population is required to confirm their equivalence. mSASHA myocardial $T_2$ values in this study (36.7±1.1 ms) were similar to previously reported values at 3T with gradient echo based techniques by van Heeswijk et al. (38.5±4.5 ms) (38) and Yang et al. (37.7±2.0 ms) (39).

Multi-parametric cardiac mapping techniques with novel and advanced reconstruction approaches have also been recently proposed to combine $T_1$ and $T_2$ mapping with cine imaging. CMR multitasking $T_1$-$T_2$ mapping combines a hybrid $T_2$IR prepared free-breathing continuous radial acquisition with a low-rank tensor model to calculate cardiac phase resolved $T_1$ and $T_2$ maps (25). CMR fingerprinting additionally uses inversion and $T_2$-preparation pulses combined with a variable flip angle pattern to induce magnetization evolutions characteristic of $T_1$/$T_2$ combinations that can be matched to a simulated dictionary (23,24). These techniques are appealing due to their free-breathing acquisition and the additional cardiac phase dimension, however are dependent on the validity of assumptions made in the modeled reconstruction. Reported in-vivo $T_1$ values from multitasking and CMR fingerprinting are similar to MOLLI (23,25), suggesting these techniques may also be susceptible to confounders.

In contrast, mSASHA uses a straightforward model with single-shot images having well-defined $T_1$ and $T_2$ weighting. Saturation pulses preceding each weighted image "reset" the magnetization to zero and remove interdependence between images as a potential confounding factor. This simplistic model has fewer assumptions and may therefore be more robust to unknown confounders. Techniques such as multitasking and fingerprinting utilize modeling of the magnetization history to calculate additional parameters, but can be subject to error when additional influences on signal evolution are not included in the model. For example, MOLLI is significantly affected by magnetization transfer (MT) because the cumulative effects of MT through each image readout is carried over to subsequent images (16). For continuous 2D implementations of multitasking and fingerprinting, through-plane motion may also present a challenging confounder to mitigate.

*The Need For Accurate Parametric Mapping*

While absolute accuracy itself is a laudable goal for parametric mapping sequences, measurement precision is often prioritized in clinical translational research. The eventual desired application of such techniques is identification of disease through abnormalities in parametric values relative to a normal reference range. Methods with higher precision narrow this range, enabling detection of more subtle deviations from a state of health, while accuracy shifts actual values for this range.

However, an understanding of the underlying mechanisms behind inaccurate measurements is important for the robustness of a method as it matures beyond small single-center studies. MOLLI's underestimation is due to a complex interaction of both physiological factors such as heart rate, $T_1$, $T_2$, and magnetization transfer, protocol parameters such as flip angle and number of phase encode lines, as well as $B_0$ and $B_1$ field inhomogeneities (15). Linear $T_2$p-bSSFP errors are primarily due to $T_1$ recovery during imaging, which is determined by the number of phase encode lines, and heart rate for long $T_1$ species. In general, confounders leading to inaccuracies are not constant and can be influenced by the sequence implementation, the operator's choice of protocol parameters, and other physiological factors. These factors are often difficult to control, particularly outside of single-center focused research studies, and reducing confidence that $T_1$/$T_2$ abnormalities reflect actual physiological changes. For sequences with known biases but unknown causes, it is even more challenging to confidently attribute measured $T_1$/$T_2$ changes to microstructural changes instead of underlying confounders.

The consensus statement from the parametric mapping working group recommends that each site establish normal values for each sequence and protocol to reduce the impact of such confounders (40). While necessary given the confounders present in many existing sequences, protocol parameters often must be changed in response to the patient's habitus or heart rate and does not fully address physiological confounders. Furthermore, this presents a significant burden for clinicians and confusion about sequence specific normative values limits the widespread adoption of parametric mapping overall.

A significant practical benefit of accurate sequences is that the measured normal values in healthy subjects are, by definition, the intrinsic $T_1$ and $T_2$ values instead of a combination of the true value and that sequence's confounding factors. Ongoing development and validation of accurate mapping sequences that are robust enough to not require site-specific normative values would significantly reduce the burden for more widespread translation of parametric mapping into clinical practice. Abnormalities in $T_1$ or $T_2$ values could also be more confidently ascribed to physiological changes instead of confounding factors.

*Limitations*

mSASHA extends upon the well validated and robust SASHA sequence and data in this study reaffirms its accuracy in phantoms with similar in-vivo $T_1$ values. Although there are no known significant confounders to the SASHA sequence, further study is required to similarly validate the mSASHA sequence. The variable flip angle readout enables the use of the 3-parmaeter model, but also results in an asymmetric modulation transfer function (MTF) across k-space. However, image sharpness in phantoms and in-vivo data did not appear to be significantly affected. Similar to many conventional techniques, mSASHA uses a single-shot readout that can be affected by cardiac motion at high heart rates and optimization of the readout and sampling strategies for higher heart rates is needed.

The in-vivo study was limited to a small cohort of healthy subjects and additional studies in a more diverse population of patients is required to assess its robustness in a more challenging clinical environment. The study was also limited to non-contrast imaging and further comparisons in subjects post-gadolinium contrast should be performed to determine relative precision and overall performance of mSASHA in-vivo with shorter $T_1/T_2$ values. mSASHA was compared only to a limited subset of widely used $T_1$ and $T_2$ mapping techniques and comparisons to newer techniques such as multitasking and cMRF would be valuable in the future.

**Conclusions**

Multi-parametric mSASHA provides accurate and precise simultaneous $T_1$ and $T_2$ quantification in a single breath-hold, delivering voxel-registered myocardial tissue characterization with reduced scan times. Further study with mSASHA is required to establish normative values at both 1.5 and 3T and demonstrate the potential of co-registered $T_1$ and $T_2$ voxel profiles to characterize and prognosticate cardiomyopathy states.